\documentclass[10pt,conference]{IEEEtran} 
\IEEEoverridecommandlockouts
\usepackage{cite}
\usepackage{amsmath,amssymb,amsfonts}
\usepackage{algorithmic}
\usepackage{graphicx}
\usepackage{textcomp}
\usepackage{subcaption}
\usepackage{xcolor}
\usepackage{tikz}
\usepackage{booktabs}
\usepackage{amsmath}

\usetikzlibrary{shapes.geometric, arrows, positioning, decorations.pathreplacing}

\usetikzlibrary{shapes.geometric, arrows}
\usepackage{algorithm}
\usepackage{algorithmic}

\definecolor{amber}{rgb}{1.0, 0.75, 0.0}
\definecolor{almond}{rgb}{0.94, 0.87, 0.8}
\definecolor{blond}{rgb}{0.98, 0.94, 0.75}
\definecolor{cornflowerblue}{rgb}{0.39, 0.58, 0.93}
\definecolor{lavenderblue}{rgb}{0.8, 0.8, 1.0}
\definecolor{lightskyblue}{rgb}{0.53, 0.81, 0.98}
\definecolor{lime(web)(x11green)}{rgb}{0.0, 1.0, 0.0}
\definecolor{lime(colorwheel)}{rgb}{0.75, 1.0, 0.0}
\definecolor{persianpink}{rgb}{0.97, 0.5, 0.75}
\definecolor{mistyrose}{rgb}{1.0, 0.89, 0.88}
\definecolor{ticklemepink}{rgb}{0.99, 0.54, 0.67}
\definecolor{salmonpink}{rgb}{1.0, 0.57, 0.64}
\definecolor{richbrilliantlavender}{rgb}{0.95, 0.65, 1.0}
\definecolor{pink}{rgb}{1.0, 0.75, 0.8}
\definecolor{cadetgrey}{rgb}{0.57, 0.64, 0.69}
\definecolor{darkpastelblue}{rgb}{0.47, 0.62, 0.8}

\def\BibTeX{{\rm B\kern-.05em{\sc i\kern-.025em b}\kern-.08em
    T\kern-.1667em\lower.7ex\hbox{E}\kern-.125emX}}
\begin{document}

\title{Performance Evaluation of Dual RIS-Assisted Received Space Shift Keying Modulation}
\author{Ferhat Bayar, Haci Ilhan, \IEEEmembership{Senior Member, IEEE} and Erdogan Aydin 
 \thanks{F. Bayar is with the Scientific and Technological Research Council of Turkey (TUBITAK), B{I}LGEM, Kocaeli, Turkey (e-mail: ferhat.bayar@tubitak.gov.tr) (Corresponding author: Ferhat Bayar).}
 \thanks{H. Ilhan is with the Department of Electronics and Communications Engineering, Yildiz Technical University, Istanbul, 34220, Turkey, (e-mail: ilhanh@yildiz.edu.tr).}
 \thanks{E. Aydin is with the Department of Electrical and Electronics Engineering, Istanbul Medeniyet University, Istanbul 34857, Turkey (e-mail: erdogan.aydin@medeniyet.edu.tr).}

}

\maketitle

\begin{abstract}
Reconfigurable intelligent surfaces (RISs) are gaining traction for their ability to reshape wireless environments with low energy consumption. However, prior studies primarily explore single-RIS deployments with static or semi-static reflection control. In this paper, we propose a novel dual-RIS-assisted architecture for smart indoor wireless signal routing, wherein the second RIS (RIS$_2$) is dynamically configured based on source data bits to steer signals toward specific receivers or indoor zones. The first RIS (RIS$_1$), positioned near a fed antenna or access point, passively reflects the incident signal. RIS$_2$, equipped with a lightweight controller, performs bit-driven spatial modulation to enable data-dependent direction selection at the physical layer. We develop a complete end-to-end system model, including multi-hop channel representation, RIS phase configuration mapping, and signal detection based on space shift keying (SSK). Performance analysis is evaluated in terms of achievable capacity and outage probability under varying inter-RIS distances and carrier frequencies. 
\end{abstract}
\begin{IEEEkeywords}
Reconfigurable intelligent surface (RIS), selective combining (SC), path loss, outage probability, multi-hop channels
\end{IEEEkeywords}

\section{Introduction}
The rapid proliferation of wireless devices in indoor environments ranging from smartphones and laptops to Internet of Things (IoT) sensors and medical monitoring equipment has created an urgent need for intelligent, flexible, and energy-efficient communication infrastructures. Conventional wireless fidelity (Wi-Fi) and cellular systems often struggle to maintain consistent coverage and performance in complex indoor layouts due to obstacles such as walls, furniture, and user movement. Reconfigurable intelligent surface (RIS), which enable the passive control of electromagnetic wave propagation, has recently been proposed as a low-cost and energy-efficient solution to enhance wireless links without the need for active relaying or expensive infrastructure upgrades \cite{liu2021reconfigurable,basar2019wireless,pan2021reconfigurable,huang2019reconfigurable }.

While existing RIS-based systems typically involve a single RIS element reflecting signals toward static or semi-dynamically selected users, such approaches remain limited in their ability to support fine-grained, data-aware control over signal direction—particularly in environments with multiple rooms, users, or zones requiring differentiated service \cite{ibrahim2021exact}. Moreover, current designs generally treat RIS control as independent of the transmitted data itself, missing opportunities for joint signal and control co-design. Recent studies have explored multi-RIS-enabled configurations, which present more practical applications for RIS-assisted networks, aiming to enhance system capacity \cite{do2021multi, katsanos2024multi,mei2020cooperative,xie2022downlink,tyrovolas2022performance, han2020cooperative}.

In this work, we propose a novel dual-RIS system architecture that addresses these limitations by introducing bit-controlled intelligent signal routing. The system consists of two RIS panels: a passive RIS$_1$ located near the transmitter or access point, and a configurable RIS$_2$ placed at a central indoor location such as a hallway wall or ceiling. Unlike traditional systems, RIS$_2$ is controlled using the data bits from the source, enabling real-time selection of the desired receiver zone or antenna by adjusting its phase profile accordingly. This allows the RIS to function as a low-energy spatial switch, dynamically guiding wireless signals toward the correct user or device based on the intended data stream.


To address these limitations, we explore the potential of RIS in enhancing wireless communication systems. Specifically, we propose a novel dual-RIS-assisted architecture for smart indoor signal routing, where the second RIS (RIS${_2}$) is dynamically configured based on source data bits to guide signals to specific receivers or indoor zones. The first RIS (RIS${_1}$) passively reflects the signal from a fed antenna or access point. Additionally, we examine the integration of RIS with space shift keying (SSK) and develop a comprehensive end-to-end system model, incorporating multi-hop channel representation and RIS phase configuration mapping. We also evaluate the system's performance in terms of achievable capacity, outage probability, and the impact of varying distances and carrier frequencies. Finally, we present simulation results that demonstrate the superior performance of the dual-RIS system compared to conventional single-RIS and fixed-beam configurations, offering a flexible and scalable solution for indoor coverage in energy and interference-constrained environments, such as smart buildings and post-disaster scenarios.

\subsection{Notations}
In this paper , matrices and vectors are shown in boldface uppercase and boldface lowercase letter, respectively. $\left(\cdot\right)^*$, $\left(\cdot\right)^T$, and $\left(\cdot\right)^H$ define complex conjugation, transpose, and Hermitian transpose, respectively.   $P_r\left[\cdot\right]$ defines the the probability of an event and $\mathbb{E}\left[X\right](\mu_X)$ denotes the mean of random variable \textit{X}.

\subsection{Paper Organization}
The remainder of this paper is organized as follows. Section II introduces the proposed dual-RIS-assisted architecture, detailing the system model, including the multi-hop channel representation and RIS phase configuration. Section III presents the performance analysis, discussing the achievable capacity and outage probability. Section IV presents the theoretical and simulation results. Finally, Section V concludes the paper with a summary of the findings and future research directions.

\begin{figure}[t]
\centering{\includegraphics[scale=0.32]{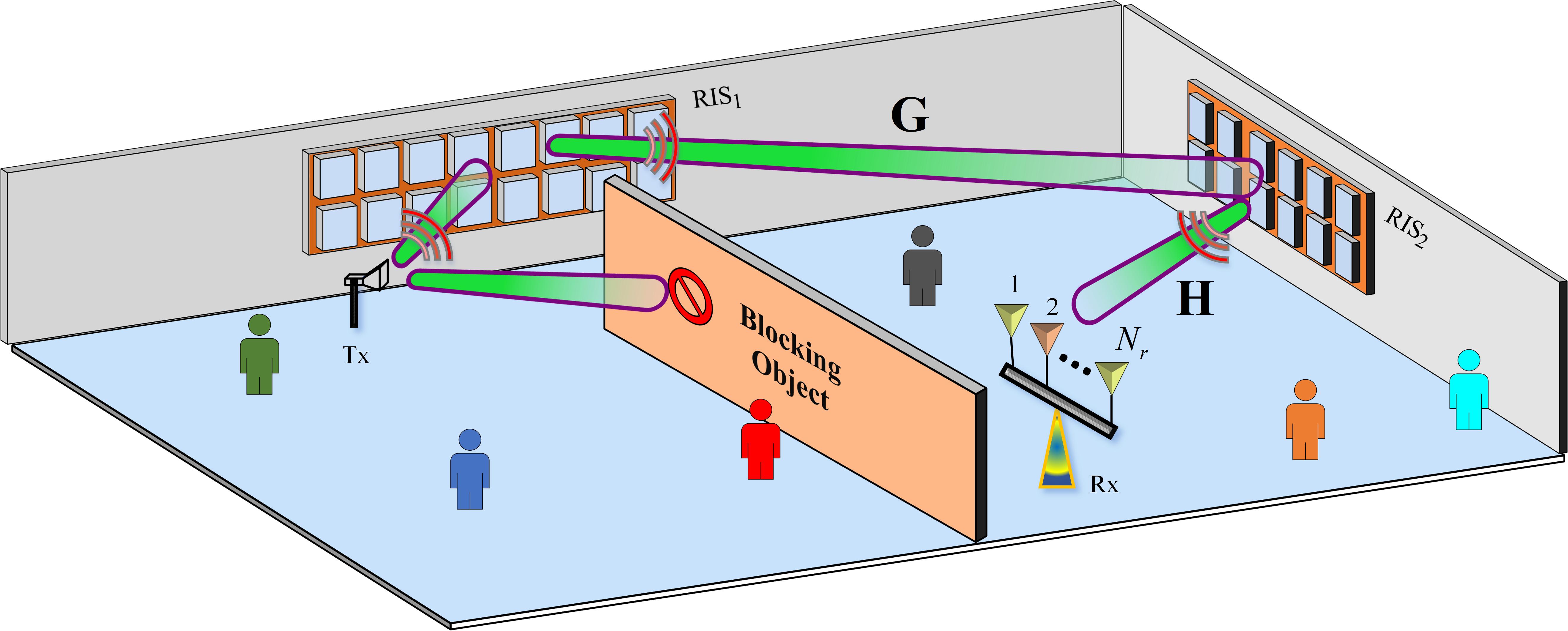}}
	\caption{Dual-RIS-assisted indoor communication system.}
	\label{p1} 		
\end{figure}

\section{System Model}
Fig.~\ref{p1} illustrates the proposed dual-RIS-assisted indoor communication system. The setup comprises a single-antenna transmitter located near a RIS$_1$, RIS$_2$ positioned in the region of the receiver, and a receiver equipped with $N_r$ antennas. The system is designed for indoor communication scenarios, such as inter-room backhaul connections, where no direct line-of-sight (LoS) path exists between the transmitter and the receiver. However, the two RISs are assumed to have direct visibility, i.e., there is an LoS link between RIS$_1$ and RIS$_2$.

Both RIS$_1$ and RIS$_2$ are equipped with $N$ passive reflecting elements each. The transmitter emits a baseband-modulated signal that is reflected by RIS$_1$, then propagates to RIS$_2$, which further reflects it toward a selected receive antenna based on SSK principles. Specifically, RIS$_2$ performs antenna selection by adjusting its phase profile such that the signal is constructively directed to one of the $N_r$ receive antennas, as determined by incoming information bits.

The system employs SSK modulation through RIS control, where $\log_2(N_r)$ bits are encoded by selecting one of the receiver antennas. RIS$_2$ uses these bits to dynamically steer the reflected signal beam toward the selected receive antenna. The system does not require any RF chains at the RIS elements, relying instead on passive beamforming to achieve efficient modulation.

\subsection{Channel Model}
The end-to-end communication channel consists of two segments. $\mathbf{G} \in \mathbb{C}^{N \times N}$ represents the channel between RIS$_1$ and RIS$_2$. Similarly, $\mathbf{H} \in \mathbb{C}^{N_r \times N}$ denotes the channel between RIS$_2$ and the receiver.

Each of these channels is modeled using a Rician fading distribution with large-scale path loss. Specifically, the wireless channel $\mathbf{G}$ and $\mathbf{H}$ can be represented as follows:
\begin{equation}
    \mathbf{G} = \begin{bmatrix}
       g_{1,1}  & \cdots & g_{1,N} 
       \\
       \vdots &  \ddots & \vdots \\ 
       g_{N,1} & \cdots & g_{N,N}
    \end{bmatrix},    \mathbf{H} = \begin{bmatrix}
       h_{1,1} & \hdots  & h_{1,N} 
       \\
       \vdots & \ddots & \vdots \\ 
       h_{N_r,1} & \hdots & h_{N_r,N}
    \end{bmatrix}
    \label{channel_matrix_bs_to_ris}
\end{equation}
where $ g_{k,i}=\alpha_{k,i}e^{-j \psi_{k,i}} \quad k= 1, 2, \cdots, \textit N, \quad i= 1, 2, \cdots, \textit N$. In here
\( \alpha_{k,i} \) and \( \psi_{k,i} \) represent the amplitude and the channel phase induced by the $k$-th reflector at the $i$-th  element of the RIS$_2$ respectively. Specifically, the envelopes of the first hop channels $\alpha_{k,i}$ (from the RIS$_{1}$ to RIS$_{2}$) are independent and identically distributed (i.i.d.) Rician random variables with ratio of the LoS component with \( K_1 \) and power \( \Omega_1 \). Similarly the the channel coefficient between the \(p\)-th received antenna and the \(i\)-th reflector element \(h_{i,p}\) is characterized as   $h_{i,p} = \beta_{i,p}e^{-j \theta_{i,p}}  \quad p \in \{ 1, \dots, N_r \}$. Here \( \theta_{i,p} \) is the channel phase induced by the $i$th reflector at the $p$-th receive antenna and $\beta_{p,i}$ defines the i.i.d. Rician random variables with ratio of the LoS component \( K_2 \) and power \( \Omega_2 \).

Since the selection of the $m$-th receiving antenna is based on the incoming bit data sequence, the channel between the (RIS$_2$) and the destination will be represented in the following form $    \mathbf{h}_{m} = \begin{bmatrix}
       h_{1,m}, & \cdots, & h_{N,m} 
    \end{bmatrix}$.
Furthermore, we assume that the channel coefficients \( h_{i,p} \) and \( g_{k,i} \) are independent. The Rician \( K \)-factors, which characterize the ratio of the power in the LoS component to that in the  non LOS (NLoS) component, are given by $K_1 = {\Omega_1}/{\sigma_1^2}$ and $K_2 = {\Omega_2}/{\sigma_2^2}$ where \( \Omega_1 \) and \( \Omega_2 \) represent the power in the LoS components for the channels \( \mathbf{G} \) and \( \mathbf{H} \), respectively, and \( \sigma_1^2 \) (for \( g_{k,i} \)) and \( \sigma_2^2 \) (for \( h_{i,p} \)) represent the powers in the NLoS components for the respective channels.

The channel coefficients \( \alpha_{k,i} \) and \( \beta_{i,p} \) are assumed to follow the Rician distribution with parameters \( K_1 \) and \( K_2 \) for the first and second hops, respectively. Specifically, the distribution of the channel coefficients is given by
\begin{equation}
    \alpha_{k,i}, \beta_{i,p} = \sqrt{{K_j}/({K_j+1})} + \sqrt{{1}/({K_j+1})} \tilde{w}
\end{equation}
where \( K_j \) represents the Rician \( K \)-factor, with \( j \in \{1, 2\} \) corresponding to the first and second hops. Here, \( \tilde{g} \) is a zero-mean complex Gaussian random variable with unit variance, i.e., \( \tilde{w} \sim \mathcal{CN}(0,1) \).

\subsection{Phase Optimization}
We assume that both RISs have access to perfect channel state information (CSI) for their respective links. The goal of the RIS phase design is to maximize the signal-to-noise ratio (SNR) at the receiver by coherently combining the reflected signals. This is achieved by optimally adjusting the phase shifts of the RIS elements to compensate for the channel-induced phase variations. Let the phase shift vector for the overall system be represented as $\boldsymbol{\Theta}_i = \left[ e^{j\phi_{i,1}}, e^{j\phi_{i,2}}, \dots, e^{j\phi_{i,N}} \right]^{\mathrm{T}},$
where \( \phi_{i,k} \) is the phase shift at the \( k \)-th RIS element. The goal is to optimize these phase shifts to maximize the received SNR. The optimization problem can be written as $\boldsymbol{\Theta}_i^{\text{opt}} = \arg \max_{\boldsymbol{\theta}_i} \left| \mathbf{h}_{\text{eff}} \right|^2, $
where \( \mathbf{h}_{\text{eff}} \) is the effective channel matrix, given by the product of the RIS$_1$-to-RIS$_2$ channel \( \mathbf{G} \) and the RIS$_2$-to-destination channel \( \mathbf{h}_m^{\mathrm{T}} \) as
$\mathbf{h}_{\text{eff}} = \mathbf{G} \mathbf{h}_m^{\mathrm{T}}.$
To maximize the received signal strength, the phase shifts must align the received signal components. The optimal phase shift for each RIS element is $\phi_{i}^{\text{opt}} = - \left( \psi_{k,i} + \theta_{i,m} \right), \quad \forall i \in \{ 1, \dots, N \},$
where \( \psi_{k,i} \) is the phase of the RIS$_1$-to-RIS$_2$ channel and \( \theta_{i,m} \) is the phase of the RIS$_2$-to-destination channel.

\subsection{Received Signal Model}
Assuming the received signal is related to the transmitted signal through the effective channel, the received signal at the \( m^{\text{th}} \) antenna can be expressed as $y_m = \sqrt{E_s} \zeta\mathbf{h}_m \mathbf{G} \boldsymbol{\Theta}_i^{\text{opt}} + n_{m},$
where $E_s$ is the square root of the transmitted signal power (\( E_s = 1 \) due to the use of the received SSK technique), \( n_m \) is the complex additive white Gaussian noise (AWGN) with zero mean and variance \( N_0 \) at the \( m \)-th receive antenna and $\zeta=\left( {\lambda_c}/{4 \pi d_1d_2} \right)^2 $. Additionally, \( d_1 \) and \( d_2 \) represent the distances from the transmitter (RIS$_1$) to RIS$_2$ and from the RIS$_2$ to the receiver, respectively. The wavelength corresponding to the center frequency \( f_c \) is given by \( \lambda_c = {c}/{f_c} \), where \( c \) is the speed of light.

The instantaneous SNR at the $m$-th selected receive antenna is then given by $    \gamma_m = {|y_m|^2}/{N_{0}} $ where \( y_m \) is the received signal at the \( m^{\text{th}} \) antenna and \( N_0 \) is the noise power. Finally, the instantaneous SNR can be expressed as
\begin{equation}
\begin{aligned}
\gamma_m =\frac{\zeta^2 E_s\left| \sum_{k=1}^{N} \sum_{i=1}^{N} \alpha_{k,i} \beta_{i,m} e^{j(\Phi^{opt}_i - \Psi_{k,i} - \theta_{i,m})} \right|^2}{N_0},
\end{aligned}
\label{eq:SNR}
\end{equation}
or equivalently $\gamma_m = {\zeta^2E_s\left| \sum_{k=1}^{N} \sum_{i=1}^{N} \alpha_{k,i} \beta_{i,m} \right|^2}/{N_0}.  $

\section{Theoretical Analaysis}

\subsection{Outage Probabilty}
In this section, we analyze the outage probability of the RIS-assisted system under Rician fading channels.  The outage probability of the system is defined as the probability that the end-to-end SNR, denoted by $\gamma_m$, falls below a predetermined outage threshold value $\gamma_{\text{out}}$, i.e., $  P_{\text{out}} = \Pr[\gamma_m \leq \gamma_{\text{out}}].$

To derive the outage probability, we first consider the distribution of $\gamma_m$. As the number of terms $N$ increases, we apply the central limit theorem (CLT), which states that the sum of a large number of independent random variables tends to a Gaussian distribution.
Although the product of two normally distributed variables is generally not itself normally distributed, it is known that under certain conditions, particularly in the asymptotic regime, the distribution of the product can be approximated as normal \cite{seijas2012approach}. Specifically, the moment-generating function (MGF) of the product of two independent normal random variables converges to that of a normal distribution as the variances tend to zero or their means grow large. In particular, if \( X \sim \mathcal{N}(\mu_X, \sigma_X^2) \) and \( Y \sim \mathcal{N}(\mu_Y, \sigma_Y^2) \), then for large \(\mu_X\), \(\mu_Y\), the product \( Z = XY \) tends toward a distribution that is approximately $  Z \sim \mathcal{N}(\mu_X \mu_Y, \mu_X^2 \sigma_Y^2 + \mu_Y^2 \sigma_X^2). $

In our case, specifically, for large $N$ the sum $\sum_{k=1}^{N} \sum_{i=1}^{N} \alpha_{k,i} \beta_{i,m}$ can be approximated as product of two Gaussian random variable with $A= \sum_{i=1}^{N}\alpha_{k,i}\sim \mathcal{N}(\mu_{\alpha},\sigma^2_{\alpha})$ and $B=\sum_{i=1}^{N}\beta_{i,m}\sim \mathcal{N}(\mu_{\beta},\sigma^2_{\beta})$. The mean of $\alpha_{k,i}$ and $\beta_{i,p}$ with Rician \(K\)-factor and total power \(\Omega\) is given by
\begin{equation}
\mu_{\alpha} = N\sqrt{{(\pi \, \Omega_1)}/({4(K_1 + 1)})} L_{1/2}(-K_1),
\end{equation}
\begin{equation}
\mu_{\beta} = N\sqrt{{(\pi \, \Omega_2)}/({4(K_2 + 1)})} L_{1/2}(-K_2),
\end{equation}
where \(L_{1/2}(-K_j)\) is the generalized Laguerre polynomial of order \(1/2\), defined as:
\begin{equation}
\begin{aligned}
L_{1/2}(-K_j) &= \exp({-K_j/2}) \left[ (1 + K_j) I_0\left({K_j}/{2}\right)\right. 
\\&+ \left. K_j I_1\left({K_j}/{2}\right) \right]
\quad j \in \{1, 2\},
\end{aligned}
\end{equation}
and \(I_0(\cdot)\), \(I_1(\cdot)\) are the modified Bessel functions of the first kind. The variance of $\alpha_{k,i}$ and $\beta_{i,p}$  can be represented as $\sigma^2_{\alpha} = N(\Omega_1 - \mu_{\alpha}^2),$ and $\sigma^2_{\beta} = N(\Omega_2 - \mu_{\beta}^2)$.  So the product $Z=AB$ is approximately distributed as
$Z \sim \mathcal{N}(\mu_\alpha \mu_\beta, \mu_\alpha^2 \sigma_\beta^2 + \mu_\beta^2 \sigma_\alpha^2)$ in the large-\(N\) limit.

Consequently, $\gamma_m$ can be approximated as a ratio of a squared Gaussian random variable and the noise power $N_0$. For large values of $N$, the distribution of $\gamma_m$ is approximately non-central Chi-square random variable with one degree of freedom, $n = 1$, and its probability density function (PDF) is given by $\gamma_m = {\zeta^2 E_s |Z|^2}/{N_0}$.
Let us define the normalized variable $X = {Z}/{\sigma_Z} \sim \mathcal{N}\!\left( {\mu_Z}/{\sigma_Z},\, 1 \right),
$ so that ${|Z|^2}/{\sigma_Z^2} = X^2$.
Since \( X^2 \) follows a noncentral chi-squared distribution with 
one degree of freedom and noncentrality parameter $ \lambda = \left( {\mu_Z}/{\sigma_Z} \right)^2 $, we have $
|Z|^2 \sim \sigma_Z^2 \, \chi^2_1(\lambda).$ 
Hence, the random variable \( \gamma_m \) can be expressed as a scaled 
noncentral chi-squared random variable as $  \gamma_m = \, \chi^2_1\!\left( {\mu_Z^2}/{\sigma_Z^2} \right){\zeta^2 E_s \sigma_Z^2}/{N_0} $ The corresponding pdf of \( \gamma_m \) is given by
\begin{equation}
\label{eq:gamma_pdf}
    f_{\gamma_m}(\gamma) 
    = \frac{1}{2 s \sigma_Z^2} 
      \exp\!\left( -\frac{\gamma/s + \mu_Z^2}{2\sigma_Z^2} \right)
      I_0\!\left( \frac{\mu_Z}{\sigma_Z^2} \sqrt{\frac{\gamma}{s}} \right),
      \quad \gamma \ge 0,
\end{equation}
where \( s = {\zeta^2 E_s}/{N_0} \). The cumulative distribution function (CDF) of $\gamma_m$, denoted by $F_{\gamma_m}(\gamma)$, represents the probability that $\gamma_m$ takes a value less than or equal to $\gamma$ can be expressed in closed form using the 
the identity $Q_1(a,b) = Q(b-a) - Q(b+a)$, the CDF $\gamma_m$ can be rewritten in terms of the first-order 
Marcum--$Q_1(\cdot)$ as
\begin{equation}
F_{\gamma_m}(\gamma)
= 1 - Q_1\!\left(
\sqrt{\frac{\mu_Z^{2}}{\sigma_Z^{2}\zeta^{2}}},\;
\sqrt{\frac{\gamma}{\sigma_Z^{2}\zeta^{2}\gamma_m}}
\right).
\end{equation}
Therefore, the outage probability is obtained as

\begin{equation}
F_{\gamma_m}(\gamma) = 1 - \left[ Q\left(z - \sqrt{\frac{\mu^2_{Z}}{\sigma_{Z}^2\zeta^2 }}\right) - Q\left(z + \sqrt{\frac{\mu^2_{Z}}{\sigma_{Z}^2\zeta^2 }}\right) \right],
\end{equation} where $z = \sqrt{ { \gamma }/{ \sigma_{Z}^2 \zeta^2 \gamma_{m} } }.$
Therefore, the outage probability $P_{\text{out}}$ can be expressed in terms of the CDF of $\gamma_m$ as $P_{\text{out}}  = F_{\gamma_m}(\gamma_{\text{out}})$. This expression provides a semi-closed-form approximation for the outage probability in RIS-assisted systems under Rician fading, incorporating both fading and path loss effects.



\subsection{Capacity Analysis}
The theoretical ergodic capacity of the RIS-assisted link can be analyzed by 
approximating the cascaded channel gain with its deterministic equivalent. 
Assuming single-stream transmission and under Rican fading, 
the instantaneous SNR is denoted by 
$\gamma_m = {\zeta^2 E_s |Z|^2}/{N_0}$, where $Z$ represents in the previous section. 

The instantaneous capacity (in bits per second per Hertz) is then given by $ C(\gamma_m) = \log_2\!\left(1 + \gamma_m \right).$ The ergodic (average) capacity is defined as $    C_{\text{avg}} = 
    \mathbb{E}_{\gamma_m}\!\left[
        \log_2\!\left(1 + \gamma_m \right)
    \right],$ where the expectation is taken with respect to $f_{\gamma_m}(\gamma)$ in~\eqref{eq:gamma_pdf}. Therefore, the average capacity is given by the following integral:
\begin{equation}
C_{\text{avg}} = \int_0^\infty \log_2(1 + \gamma_m) f_{\gamma_m}(\gamma_m) \, d\gamma_m.   
\label{avarage_capacity_int}
\end{equation}

Since obtaining a closed-form expression for~\eqref{avarage_capacity_int} is 
generally intractable, a tractable upper-bound or deterministic-equivalent 
approximation can be used. In the large-$N$ regime (\( N \gg 1 \)), where the 
law of large numbers ensures that random fluctuations of $|Z|^2$ vanish, 
the instantaneous SNR $\gamma_m$ converges to its mean value 
$\bar{\gamma}_m = \mathbb{E}[\gamma_m] = {\zeta^2 E_s \mu_Z^2}/ {N_0}$.
Hence, the ergodic capacity can be approximated as
\begin{equation}
\label{eq:C_avg_final}
    C_{\text{avg}} \approx 
    \log_2\!\left(1 + \bar{\gamma}_m \right)
    = \log_2\!\left(1 + \frac{\zeta^2 E_s \mu_Z^2}{N_0} \right),
\end{equation}
which provides an upper-bound estimate of the achievable rate when the RIS 
elements are phase-aligned to coherently combine the reflected signals. 
This upper bound equivalent is particularly accurate in the 
high-RIS-element regime, where channel hardening occurs and the system 
performance approaches that of a deterministic channel.

\section{Simulation Results}
\begin{figure}[htb]
     \centering
     \includegraphics[width=0.49\textwidth]{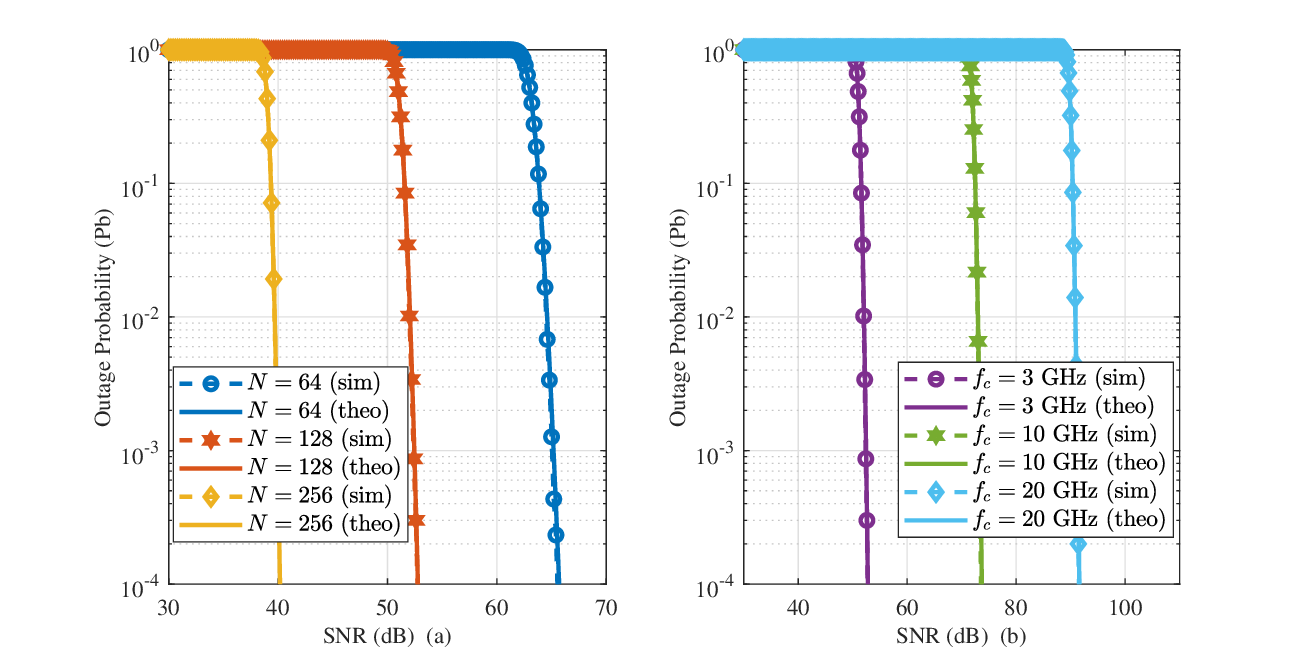}
     \caption{Outage performance of the Dual-RIS assisted RSSK schemes: (a)  with different $N$, (b) with different $f_c$ ($N_r=2$, $f_c= 3$GHz,  $K=2$, $d=10$, and $\gamma_{\text{out}} = 10$ dB).}
     \label{Fig1}
   \end{figure}

In Fig.\ref{Fig1} (a), the effect of varying the number of reflecting elements on the outage probability is represented. The figure compares three different values of the number of reflecting elements: $N = 64, 128, 256$, while keeping the operating frequency fixed at $f_c = 3 \text{GHz}$, the distance at $d=10$ m, the number of receive antennas $N_r=2$ and $\gamma_{\text{out}} = 10$ dB. The SNR parameter used in the simulations  is expressed as $\mathrm{SNR (dB)}=10\log_{10}(E_s/N_0)$. As observed in the Fig.\ref{Fig1} (a), the outage probability decreases significantly as the number of reflecting elements increases. This suggests that the system's performance improves with the deployment of a greater number of reflecting elements. Specifically, for $N = 64$, the outage probability is relatively higher, indicating a lower reliability of the communication link. However, as $N$ is increased to $128$, a noticeable reduction in the outage probability is observed, which becomes even more pronounced for $N = 256$. The latter configuration, with $256$ reflecting elements, results in the lowest outage probability across the entire range of SNRs. This behavior can be attributed to the increased ability of the larger number of reflecting elements to enhance the channel conditions, thereby improving signal strength and reducing interference, which leads to a more reliable link.

Fig.\ref{Fig1} (b) illustrates the effect of center frequency ($f_c = 3 \text{GHz}$ and $f_c = 10 \text{GHz}$) on the outage probability of the system. The analysis is conducted under the assumption that the number of reflecting elements is fixed at $N = 64$ and the number of receive antennas $N_r =2$ for all configurations. Specifically, the frequency of 3 GHz consistently exhibits the lowest outage probability throughout the SNR. This indicates a more reliable communication link for this frequency, as it maintains a relatively low outage probability even at higher SNR values. In contrast, the frequencies of 10 GHz and 20 GHz exhibit higher outage probabilities as the SNR increases. This indicates a more severe degradation in link reliability for the higher frequency bands. The observed differences in outage probability can be attributed to the distinct propagation characteristics associated with each frequency band. Specifically, the 3 GHz frequency benefits from lower path loss and is less affected by atmospheric absorption, resulting in a more reliable communication link. 
\begin{figure}[htb]
     \centering
     \includegraphics[width=0.49\textwidth]{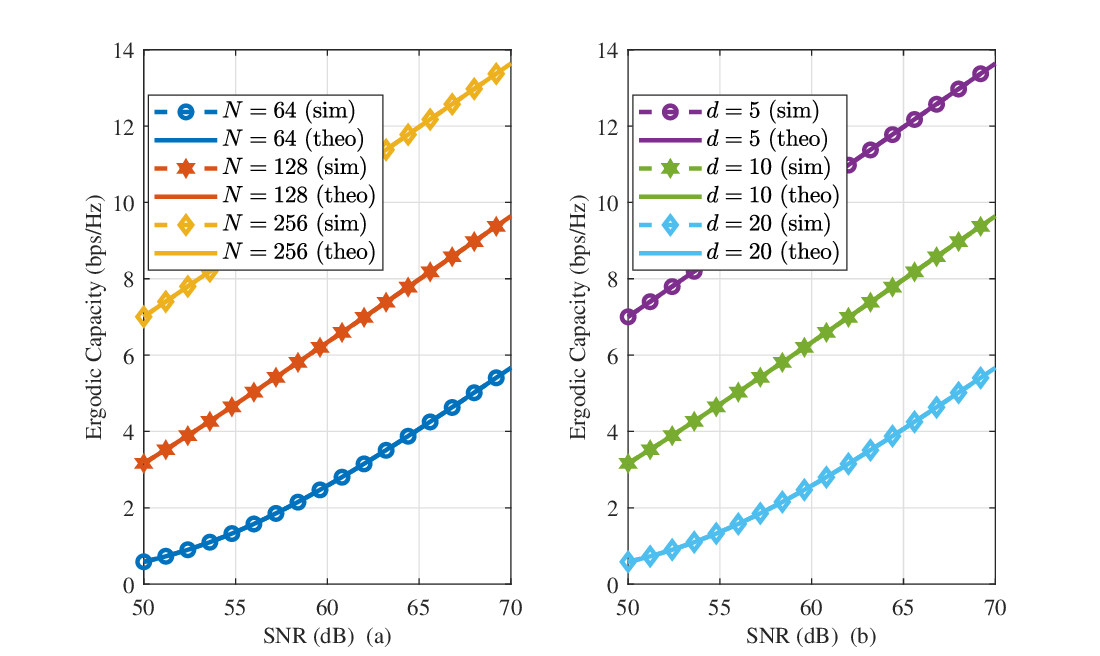}
     \caption{Ergodic capacity of the Dual-RIS assisted RSSK schemes : (a) with different number of reflecting elements $N$, (b) with different distances $d$.  ($N=128$, $N_r=2$, $f_c=3$GHz.}
    \label{Fig2}
\end{figure}
Fig.\ref{Fig2} (a) illustrates the effect of number of reflecting elements ($N = 64, 128, 256$ ) on the ergodic capacity of the system while keeping the operating frequency fixed at $f_c = 3 \text{GHz}$, the distance at $d=10$ m, the number of receive antennas $N_r=2$. As observed in the Fig.\ref{Fig2} (a), the ergodic capacity improves significantly as the number of reflecting elements increases. Fig.\ref{Fig2} (b) illustrates the effect of distance ($d = 5, 10, 20$ m) on the ergodic capacity of the system. The analysis is conducted under the assumption that the number of reflecting elements is fixed at $N = 128$ for all systems. Furthermore, the number of receive antennas, $N_r$, is set to $2$ for all configurations, and the center frequency is selected as $f_c = 3 \text{GHz}$. As the distance between the transmitter and receiver increases, a noticeable degradation in ergodic capacity is observed. This decline can be attributed to the increase in path loss with distance, which results in reduced signal strength and consequently lower achievable data rates. At a distance of \(d = 5\) m, the system experiences minimal attenuation, yielding the highest ergodic capacity. However, as the distance increases to \(d = 10\) m and \(d = 20\) m, the capacity diminishes due to greater signal propagation loss over longer distances.



\begin{figure}[htb]
\centering
\includegraphics[width=0.49\textwidth]{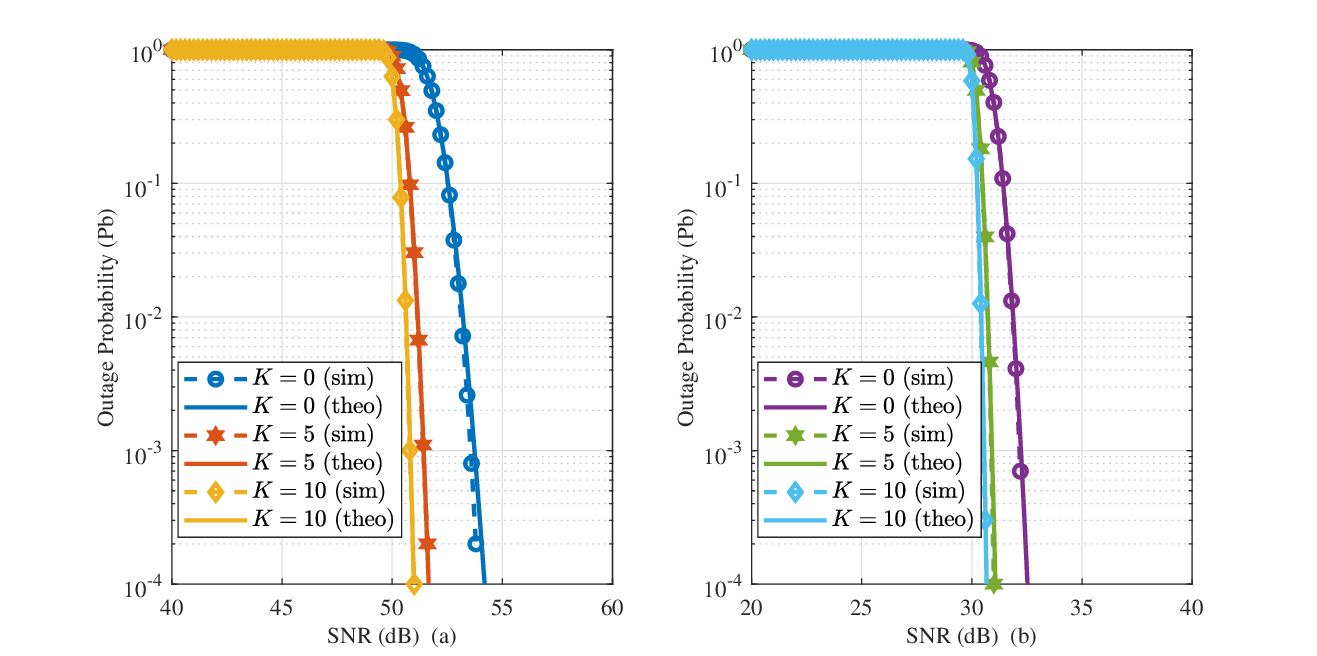}
\caption{Outage performance of the RIS assisted RSSK systems: (a) Dual-RIS with different $K$, (b) Single-RIS with different $K$ ($N=128$, $N_r=2$, $d=10$ m, $f_c=3$ GHz and $\gamma_{\text{out}} = 10$ dB).}
\label{Fig4}
\end{figure}


Fig.\ref{Fig4} illustrates the effect of the dual-RIS and single-RIS configurations on the outage probability of the system. The analysis is conducted under the assumption that the number of reflecting elements is fixed at $N = 128$ and the number of receive antennas $N_r = 2$ for all configurations. Specifically, the single RIS configuration consistently exhibits the lowest outage probability across all values of $K$ throughout the SNR range as seen in Fig.\ref{Fig4} (b). This indicates a more reliable communication link for the single RIS system, as it maintains a relatively low outage probability even as $K$ increases and the SNR becomes higher. In contrast, the dual-RIS configuration shows a higher outage probability compared to single RIS in all instances, even as $K$ increases as shown in Fig.\ref{Fig4} (a). This suggests that the additional RIS element (RIS$_2$) introduces complexity that affects the system’s ability to minimize outages effectively. Despite the potential for enhanced signal routing and coverage flexibility offered by dual RIS, it does not outperform single RIS in terms of outage probability under the evaluated conditions. The observed differences in outage probability can be attributed to the complexity associated with managing multiple RIS surfaces in the dual-RIS setup. While dual RIS improves coverage and signal routing flexibility, the coordination overhead and interference effects may counteract the expected benefits, leading to a higher outage probability compared to the simpler, more efficient single RIS system.

\section{Conclusions and Future Works}

In conclusion, the simulation results highlight the potential of the dual-RIS system in providing scalable, flexible, and energy-efficient solutions for indoor coverage, particularly in challenging environments like smart buildings and post-disaster scenarios. The proposed approach not only meets the growing demands of five generation (5G) and six generation (6G) networks but also lays the groundwork for future research on multi-RIS systems and their integration into next-generation wireless communication frameworks.The dual-RIS-assisted architecture presents a promising direction for future wireless communication systems, offering substantial improvements in signal routing, coverage, and interference mitigation. Future work will focus on further optimizing the system model, exploring real-world implementation challenges, and extending the proposed framework to support emerging technologies in wireless communication.

\vspace{12pt}

\bibliographystyle{IEEEtran}
\bibliography{IEEEabrv}

\end{document}